\documentclass[12pt]{article}
\usepackage{amssymb}
\usepackage{epsf}
\usepackage{graphicx}

\def\beq{\begin{equation}}
\def\eeq{\end{equation}}
\def\bea{\begin{eqnarray}}
\def\eea{\end{eqnarray}}
\def\ba{\begin{array}}
\def\ea{\end{array}}
\def\bea{\begin{eqnarray}}
\def\eea{\end{eqnarray}}

\begin{document}
\title{Exact Solutions of Schr\"{o}dinger Equation for a Ring-Shaped Potential}
\author{\vspace{1cm}
         {\"{O}zlem Ye\c{s}ilta\c{s}$^1$, Ramazan Sever $^2$}
         \thanks{Corresponding Author: sever@metu.edu.tr}
\\
{\small \sl  $^1$Gazi University,Faculty of Arts and Sciences,
Department of Physics,06500,Ankara Turkiye}
\\{\small \sl $^2$Middle East Technical University,Department of
Physics,06531 Ankara, Turkiye }}

\maketitle

\begin{abstract}
\noindent The exact solutions of Schr\"{o}dinger equation are
obtained for a noncentral potential which is a ring-shaped
potential. The energy eigenvalues and corresponding eigenfunctions
are obtained for any angular momentum $\ell$. Nikiforov-Uvarov
method is used in the computations.
\end{abstract}
~~~~~~~~{\small \sl PACS numbers: 03.65.Db, 03.65.Ge} 

\newpage

\section{Introduction}
\noindent Exact solutions of the fundamental wave equations of the
non-relativistic quantum mechanics have attracted much attention in
recent years. So far many efforts have been made to solve the
time-independent Schr\"{o}dinger equation for non-central potentials
in three and two dimensions such as the Coulombic ring-shaped
potential [1-5] and deformed ring-shaped potential [6], the Hellmann
potential [7]. In quantum mechanics,  Aharonov-Bohm (AB) effect is
an another popular non-central potential type [8]. It is also
studied through the wide areas such as path integral solutions[9],
in non-commutative quantum mechanics [10], a Dirac particle in the
Aharonov-Bohm potential [11-13] and scattering [14]. On the other
hand, AB potential is also known as ring shape potential [6].

\noindent Recently, an alternative and simple method called as
Nikiforov-Uvarov Method [15] (NU-method) has been introduced for
solving the Schr\"{o}dinger equation for some well known potentials
[16-18]. It is also used to solve the  Klein-Gordon and
Duffin-Kemmer-Petiau wave equations in the presence of exponential
type potentials such as standard Woods-Saxon [19], P\"{o}schl-Teller
[20] and Hulthen potentials [21-23]. The Nikiforov-Uvarov method
[15] is used in the computations to obtain energy eigenvalues and
the corresponding eigenfunctions. Recently Alhaidari et al have
studied Dirac and Klein Gordon equation for the Coulomb, oscillator
and Hartmann potentials [24].

Ring shaped potential has an application field in quantum chemistry
as a model for the Benzene molecule given as [1,4,9] \
\begin{eqnarray}
V_{q}(r)=\eta \sigma^{2} \epsilon_{0}\left(\frac{2a_{0}}{r}-\frac{
\eta a^{2}_{0}}{r^{2}sin^{2}\theta}\right)
\end{eqnarray}
where $a_{0}=\frac{\hbar^{2}}{\mu e^{2}}, \,\
\epsilon_{0}=-\frac{1}{2}\frac{\mu e^{4}}{\hbar^{2}}$, Bohr radius
and the ground state energy of the hydrogen atom, respectively,
$\mu$ is the particle mass, $\eta$ and $\sigma$ are dimensionless
positive real parameters which range from about $1$ to $10$ [9].

The paper is arranged as follows. In section II, the
Nikiforov-Uvarov Method is given briefly. In section III, NU method
is applied to three dimensional Schr\"{o}dinger equation with the
ring shaped potential. Results are discussed in section IV.

\section{The Nikiforov-Uvarov Method}
The NU-method developed by Nikiforov and Uvarov is based on reducing
the second order differential equations (ODEs) to a generalized
equation of hypergeometric type [15]. It is provided us an analytic
solution of Schr\"{o}dinger equation for certain kind of potentials.
This is based on the solutions of general second order linear
differential equation with orthogonal functions [15]. For a given
appropriate potential, the one-dimensional Schr\"{o}dinger equation
is reduced to a generalized equation of hypergeometric type with a
suitable coordinate transformation $s=s(r)$. Then, the equation has
the form,
\begin{eqnarray}
\psi^{''}(s)+\frac{\tilde{\tau}(s)}{\sigma(s)}\psi^{'}(s)+
\frac{\tilde{\sigma}(s)}{\sigma^{2}(s)}\psi(s)=0
\end{eqnarray}
where $A(s)=\frac{\tilde{\tau}(s)}{\sigma(s)}$ and
$B(s)=\frac{\tilde{\sigma}(s)}{\sigma^{2}(s)}$. In Eq.(2),\,\
$\sigma(s)$ and $\tilde{\sigma} (s)$ are polynomials at most second
degree. $\tilde{\tau}(s)$ is a polynomial with at most first degree
[15]. The wave function is constructed as a multiple of two
independent parts,
\begin{eqnarray}
\psi(s)=\phi(s) y(s),
\end{eqnarray}
and Eq.(2) becomes [15]
\begin{eqnarray}
\sigma(s)y^{''}(s)+\tau(s)y^{'}(s)+\lambda y(s)=0,
\end{eqnarray}
where
\begin{eqnarray}
\sigma(s)=\pi(s)\frac{d}{ds}(ln \phi(s)),
\end{eqnarray}
and
\begin{eqnarray}
\tau(s)=\tilde{\tau}(s)+2\pi(s).
\end{eqnarray}
$\lambda$ is defined as
\begin{eqnarray}
\lambda_{n}+n\tau^{'}+\frac{[n(n-1)\sigma^{''}]}{2}=0, n=0,1,2,...
\end{eqnarray}
determine $\pi(s)$ and $\lambda$ by defining
\begin{eqnarray}
k=\lambda-\pi^{'}(s).
\end{eqnarray}
and the function $\pi(s)$ becomes
\begin{eqnarray}
\pi(s)=(\frac{\sigma^{'}-\tilde{\tau}}{2})\pm
\sqrt{(\frac{\sigma^{'}-\tilde{\tau}}{2})^{2}-\tilde{\sigma}+k\sigma}
\end{eqnarray}
The polynomial $\pi(s)$ with the parameter $s$ and prime factors
show the differentials at first degree. Since $\pi(s)$ has to be a
polynomial of degree at most one, in Eq.(9) the expression under the
square root must be the square of a polynomial of first degree [15].
This is possible only if its discriminant is zero. After defining
$k$, one can obtain $\pi(s)$, $\tau(s)$,$\phi(s)$ and $\lambda$. If
we look at Eq.(5) and the Rodrigues relation
\begin{eqnarray}
y_{n}(s)=\frac{B_{n}}{\rho(s)}\frac{d^{n}}{ds^{n}}[\sigma^{n}(s)\rho(s)],
\end{eqnarray}
where $C_{n}$ is normalization constant and the weight function
satisfy the relation as
\begin{eqnarray}
\frac{d}{ds}[\sigma(s)\rho(s)]=\tau(s)\rho(s).
\end{eqnarray}
where
\begin{eqnarray}
\frac{\phi^{'}(s)}{\phi(s)}=\frac{\pi(s)}{\sigma(s)}.
\end{eqnarray}
\section{Solutions of The Ring-Shaped Potential}
Let us start with the Schr\"{o}dinger equation in three dimensions
written as
\begin{eqnarray}
-\frac{\hbar^{2}}{2\mu}
\left[\frac{1}{r^{2}}\frac{\partial}{\partial
r}\left(r^{2}\frac{\partial}{\partial
r}\right)+\frac{1}{r^{2}}\left(\frac{1}{\sin
\theta}\frac{\partial}{\partial \theta}\left(\sin \theta
\frac{\partial}{\partial\theta}\right)+
\frac{1}{\sin^{2}\theta}\frac{\partial^{2}}{\partial\varphi^{2}}\right)+
V(r,\theta)-E \right]\phi(r,\theta,\varphi)=0
\end{eqnarray}
where $V(r,\theta)$ is taken as below
\begin{eqnarray}
V(r,\theta)=-\frac{A}{r}+\frac{B}{r^{2}sin^{2}\theta}
\end{eqnarray}
$A$ and $B$ parameters are used instead of the parameters defined in
Eq.(1) for simplicity. If we take the spherical wave function as
$\phi(r,\theta,\varphi)=R(r)Y(\theta, \varphi)$ \,\ in order to
separate the  Schr\"{o}dinger equation into variables, we obtain
\begin{eqnarray}
\frac{1}{R(r)}\frac{\partial}{\partial r}\left(r^{2}\frac{\partial
R(r)}{\partial r}\right)+\frac{2\mu A r}{\hbar^{2}}+\frac{2\mu
Er^{2}}{\hbar^{2}}+\frac{1}{Y(\theta, \varphi)}\frac{1}{\sin
\theta}\frac{\partial}{\partial\theta}\left(\sin \theta
\frac{\partial Y(\theta, \varphi)}{\partial\theta}\right)-\\
\frac{2\mu B}{\hbar^{2}sin^{2}\theta}+\frac{1}{Y(\theta,
\varphi)}\frac{1}{sin^{2}\theta}\frac{\partial^{2}Y(\theta,
\varphi)}{\partial\varphi^{2}}=0
\end{eqnarray}
Using $Y(\theta, \varphi)= H(\theta) \Phi(\varphi)$ to separate the
Schr\"{o}dinger equation into variables, the following three
equations are obtained as:
\begin{eqnarray}
\frac{d^{2}R(r)}{dr^{2}}+\frac{2}{r}\frac{dR(r)}{dr}+\frac{2\mu
}{\hbar^{2}}
\left(E+\frac{A}{r}-\frac{\hbar^{2}}{2\mu}\frac{\lambda}{r^{2}}\right)R(r)=0,
\end{eqnarray}

\begin{eqnarray}
\frac{d^{2}H(\theta)}{d\theta^{2}}+cot\theta
\frac{dH(\theta)}{d\theta}+\left(\lambda-\frac{m^{2}}{sin^{2}\theta}-
\frac{2\mu B}{\hbar^{2}sin^{2}\theta}\right)H(\theta)=0,
\end{eqnarray}
and
\begin{eqnarray}
\frac{d^{2}\Phi(\varphi)}{d\varphi^{2}}+m^{2}\Phi(\varphi)=0.
\end{eqnarray}
where $m^{2}$ and $\lambda$ are separation constants. The solution
of Eq.(18) is
\begin{eqnarray}
\Phi_{m}=Ae^{im\varphi}, \,\ m=0, \pm 1, \pm 2,...
\end{eqnarray}
We will first solve the radial part of the Schr\"{o}dinger equation.
If we put $R(r)=\frac{G(r)}{r}$ which is bounded as $r\rightarrow 0$
and Eq.(16) becomes
\begin{eqnarray}
G^{''}(r)+\frac{1}{r^{2}}\left(-\varepsilon^{2}r^{2}-b^{2}r-\ell(\ell+1)\right)G(r)=0.
\end{eqnarray}
where $\varepsilon^{2}=-\frac{2\mu E}{\hbar^{2}}$, \,\
$b^{2}=-\frac{2\mu A}{\hbar^{2}}$ and $\lambda= \ell (\ell+1)$.
\begin{eqnarray}
\tilde{\tau}(r)=0,\,\ \sigma(r)=r,\,\
\tilde{\sigma}(r)=-\varepsilon^{2}r^{2}-b^{2}r-\ell (\ell+1)
\end{eqnarray}
Using Eq.(9),\,\ $\pi(r)$ is found in four possible values as
\begin{eqnarray}
\pi(r)=\frac{1}{2}\pm \left\{
  \begin{array}{ll}
    (\sqrt{\varepsilon^{2}}\,\ r+\ell+\frac{1}{2}),\,\ k_{1}=-b^{2}+
    2\sqrt{\varepsilon^{2}}(\ell+\frac{1}{2}) \hbox{;} \\
    (\sqrt{\varepsilon^{2}}\,\ r-\ell-\frac{1}{2}),\,\ k_{2}=-b^{2}-
    2\sqrt{\varepsilon^{2}}(\ell+\frac{1}{2}) \hbox{.}
  \end{array}
\right.
\end{eqnarray}
where $k_{1,2}$ is determined by means of the same procedure as in
[15]. From Eq.(6) we obtain
\begin{eqnarray}
\tau(r)=\left\{
          \begin{array}{ll}
            1+2(\sqrt{\varepsilon^{2}}r+\ell+\frac{1}{2}),\,\,\ k_{1}=-b^{2}+
    2\sqrt{\varepsilon^{2}}(\ell+\frac{1}{2}) \hbox{,} \\
            1-2(\sqrt{\varepsilon^{2}}r+\ell+\frac{1}{2}),\,\,\ k_{1}=-b^{2}+
    2\sqrt{\varepsilon^{2}}(\ell+\frac{1}{2}) \hbox{,} \\
            1+2(\sqrt{\varepsilon^{2}}r+\ell+\frac{1}{2}),\,\,\
            k_{1}=-b^{2}-
    2\sqrt{\varepsilon^{2}}(\ell+\frac{1}{2}) \hbox{,} \\
            1-2(\sqrt{\varepsilon^{2}}r-\ell-\frac{1}{2}),\,\,\ k_{2}=-b^{2}-
    2\sqrt{\varepsilon^{2}}(\ell+\frac{1}{2})  \hbox{.}
          \end{array}
        \right.
\end{eqnarray}
Imposing $\tau^{'}(r)< 0$ [15], for $k_{2}=-b^{2}-
    2\sqrt{\varepsilon^{2}}(\ell+\frac{1}{2})$ \,\ we obtain energy
    eigenvalues as
\begin{eqnarray}
E_{n_{1}, \ell}=-\frac{\mu A^{2}}{2 \hbar^{2}}
\frac{1}{(n_{1}+\ell+1)^{2}}
\end{eqnarray}
which agrees with the results [5]. Following the same steps, we find
the weight function as
\begin{eqnarray}
\rho(s)=r^{2\ell+1}e^{-\sqrt{\varepsilon^{2}}r}
\end{eqnarray}
$y_{n_{1}}$ is obtained by using Eq.(10)
\begin{eqnarray}
y_{n_{1}, \ell}(r)= L^{(2 \ell+1)}_{n_{1}}(r)
\end{eqnarray}
and also other part of the wave function
\begin{eqnarray}
\phi(r)=r^{\ell+1} e^{-\sqrt{\varepsilon^{2}}r}
\end{eqnarray}
finally, the radial part solution is obtained as
\begin{eqnarray}
G_{n_{1}, \ell}(r)=A_{n_{1}, \ell} \,\ r^{2
\ell+1}e^{-\sqrt{\varepsilon^{2}}r}L^{2\ell+1}_{n_{1}}(r)
\end{eqnarray}
where  $A_{n,\ell}$ is a normalization constant
\begin{eqnarray}
A_{n_{1}, \ell}=\sqrt{\frac{n_{1}
!}{2(n_{1}+2\ell+1)(n_{1}+2\ell+1)!}} \nonumber
\end{eqnarray}
Hence,  these radial solutions are consistent with refs.[5,17].

Now we solve angular part of the Schr\"{o}dinger equation given in
Eq.(17). Defining a new variable as
\begin{eqnarray}
x=\cos \theta
\end{eqnarray}
then, Eq.(17) takes the form
\begin{eqnarray}
\Theta^{''}(x)-\frac{2x}{1-x^{2}}\Theta^{'}+\frac{1}{1-x^{2}}^{2}\left(\lambda
(1-x^{2})-(m^{2}+\beta)\right)\Theta(x)=0
\end{eqnarray}
To apply the NU method, we define the polynomials
\begin{eqnarray}
\tilde{\tau}=-2x,\,\ \sigma=1-x^{2},\,\ \tilde{\sigma}=-\lambda
x^{2}+(\lambda-m^{2}-\beta)
\end{eqnarray}
where $\beta=\frac{2\mu B}{\hbar^{2}}$ \,\ and $\pi(x)$ becomes
\begin{eqnarray}
\pi(x)=\pm \left\{
  \begin{array}{ll}
    (\sqrt{m^{2}+\beta},\,\ k_{1}=\lambda \hbox{;} \\
    (\sqrt{(m^{2}+\beta)}\,\ x,\,\ k_{2}=\lambda-m^{2}-\beta \hbox{.}
  \end{array}
\right.
\end{eqnarray}
where $k_{1,2}$ is determined by means of the same procedure as in
[15]. We obtain $\tau$ as
\begin{eqnarray}
\tau(x)=\left\{
          \begin{array}{ll}
            -2x+2\sqrt{\beta+m^{2}},\,\  k_{1}=\lambda \hbox{,} \\
            -2x-2\sqrt{\beta+m^{2}},\,\  k_{1}=\lambda \hbox{,} \\
            -2x+2\sqrt{\beta+m^{2}}x,\,\ k_{2}=\lambda-m^{2}-\beta \hbox{,} \\
            -2x-2\sqrt{\beta+m^{2}}x,\,\ k_{2}=\lambda-m^{2}-\beta \hbox{.}
          \end{array}
        \right.
\end{eqnarray}
We impose $\tau^{'}(x)< 0$ because of the physical solutions. For
$k_{2}=\lambda-m^{2}-\beta$, we use negative sign of $\pi$ in this
case. Thus we obtain
\begin{eqnarray}
\ell=n_{2}+\sqrt{m^{2}+\beta}
\end{eqnarray}
where we use $\lambda=\ell (\ell+1)$. In order to get the wave
function of the polar angle of Schr\"{o}dinger equation, we use
Eqs.(3) and (10-12) and one obtains,
\begin{eqnarray}
\phi=(1-x^{2})^{\frac{1}{2}\sqrt{m^{2}+\beta}}
\end{eqnarray}
and
\begin{eqnarray}
\rho=(1-x^{2})^{-\sqrt{m^{2}+\beta}}
\end{eqnarray}
\begin{eqnarray}
y_{n}=B_{n}
(1-x^{2})^{\sqrt{m^{2}+\beta}}\frac{d^{n}}{dx^{n}}((1-x^{2})^{n-\sqrt{m^{2}+\beta}})
\end{eqnarray}
where $B_{n}$ is a normalization constant and the wave function is
given as
\begin{eqnarray}
\Theta_{n_{2},\ell}(\theta)=C_{n_{2}}(\sin
\theta)^{m^{'}}P^{(m^{'},m^{'})}_{n_{2}}(\cos \theta)
\end{eqnarray}
where $m^{'}=\sqrt{m^{2}+\beta}$ and $C_{n_{2}}$ is a normalization
constant. Eq.(39) agrees with the results [5]. $C_{n_{2}}$ is given
as:
\begin{eqnarray}
C_{n_{2}}=\sqrt{\frac{(2n_{2}+2m^{'}+1)\Gamma(n_{2}+1)\Gamma(n_{2}+2m^{'}+1)}{2^{2m^{'}+1}
\Gamma(2n_{2}+m^{'}+1)\Gamma(n_{2}+m^{'}+1)}} \nonumber
\end{eqnarray}
Therefore, the total energy of the system in Eq.(13) is given as (in
a.u: $\mu=e=\hbar=1,  a_{0}=1,\,\ \epsilon_{0}=-1/2$)
\begin{eqnarray}
E_{(n_{1},n_{2}),m}=-\frac{\eta^{2}\sigma^{4}}{2(\sqrt{m^{2}+\beta}+n_{1}+n_{2}+1)^{2}}
\end{eqnarray}
which compares with ref.[1].

\section{Conclusions}
We have solved  Schr\"{o}dinger equation for the  ring shaped
potential. The energy eigenvalues and the corresponding
eigenfunctions are obtained exactly by using NU-method.
Eigenfunctions are expressed in terms of Laguerre and Jacobi
polynomials for radial and angular parts respectively. It is seen
that NU method is an applicable tool for not only central potentials
but also noncentral and combined potentials. Results are in good
agreement with the earlier works [1,5,6]. Some numerical results are
given in Table 1. It is pointed out that for the values of $B=0$,
the results are in good agreement with ref. [6,25].

\section{Acknowledgements}

This research was partially supported by the Scientific and
Technological Research Council of Turkey.

\newpage
\begin{table}
        \begin{center}
        \caption{Energy levels for $\eta \sigma^{2}=Z$, $\eta=1$}
\begin{tabular}{|r|r|r|r|}
 \hline
$n_{1}$&$n_{2}$&$m$&$E$\\
\hline
1&1&0&-0.849763\\
2&2&0&-0.377661\\
& &1&-0.330452\\
3&3&2&-0.178302\\
& &1&-0.192028\\
& &0&-0.212431\\
4&4&3&-0.112357\\
& &2&-0.118038\\
& &1&-0.125353\\
& &0&-0.135954\\
5&5&4&-0.0776005\\
& &3&-0.0804445\\
& &2&-0.0838663\\
& &1&-0.0882161\\
& &0&-0.0944122\\
 \hline
\end{tabular}
\end{center}
\end{table}

\end{document}